\documentclass[aps,prb,twocolumn,groupedaddress,showpacs]{revtex4}
\usepackage{amssymb}
\usepackage[fleqn]{amsmath}
\usepackage[dvips]{graphicx}
\usepackage{docs}
\usepackage{bm}
\usepackage{textcomp}
\usepackage[colorlinks=true,linkcolor=blue,citecolor=blue,urlcolor=black]{hyperref}%
\expandafter\ifx\csname package@font\endcsname\relax\else
 \expandafter\expandafter
 \expandafter\usepackage
 \expandafter\expandafter
 \expandafter{\csname package@font\endcsname}%
\fi

\usepackage[papersize={9in,12in}]{geometry}
\newlength{\Figwidth}
\newcommand{\Fig}[1] {Fig.~\ref{#1}}

\begin{document}
\title{Spin Relaxation at Graphene Nanoribbons in the presence of Substrate Surface Roughness}

\author{Zahra~Chaghazardi{\color{blue}$^{1}$}}
\author{Shoeib~Babaee Touski{\color{blue}$^{2}$}}
\author{\\Rahim Faez{\color{blue}$^{1}$}}
\author{Mahdi Pourfath{\color{blue}$^{2,3}$}\footnote[1]{pourfath@ut.ac.ir }}
\affiliation{{\color{blue}$^{1}$}School of Electrical Engineering, Sharif University of Technology, Tehran, Iran}
\affiliation{{\color{blue}$^{2}$}School of Electrical and Computer Engineering, University of Tehran, Tehran, Iran}
\affiliation{{\color{blue}$^{3}$}Institute for Microelectronics, Technische Universit\"at Wien, Gu{\ss}hausstra{\ss}e 27--29/E360, A-1040 Wien, Austria}

\begin{abstract}
In this work spin transport in corrugated armchair graphene nanoribbons (AGNR) is studied. We survey combined effects of spin-orbit interaction and surface roughness, employing the non-equilibrium Green's function formalism and four orbitals tight-binding model. We modify hopping parameters regarding bending and distance of corrugated carbon atoms. The effects of surface roughness parameters, such as roughness amplitude and correlation length, on the spin transport of the graphene nanoribbons are studied. We show that increasing surface roughness amplitude breaks the AGNR symmetry and hybridize $\mathit{\sigma}$ and $\mathit{\pi}$ orbitals, leading to more spin flipping and therefore decrease in polarization. Unlike the roughness amplitude, the longer correlation length makes AGNR surface smoother and increases polarization. Moreover, Spin diffusion length of carriers is extracted and its dependency on the roughness parameters is investigated.  We find the spin diffusion length for various surface corrugation amplitude in order of 1 to 80 micrometer . 
\end{abstract}



\maketitle

 \section{introduction}

Graphene-- a monolayer of graphite-- is an attractive material for electronic and spintronic devices due to its unique properties.\cite{han2014graphene}. High charge mobility\cite{bolotin2008} and long spin relaxation time \cite{Guinea06} make it as an interesting base material for spintronics.
 It is supposed to have relatively small spin orbit interaction due to low atomic number of carbon $\mathrm {Z=6}$ ~\cite{Guinea06}. Based on theoretical calculations, various values are estimated for the gap opened by intrinsic spin-orbit coupling (ISOC) at the $\mathrm {K}$ point. An early study predicts ISOC gap as $ 200 \mu \mathrm{eV}$ ~\cite{kane05}. A later study calculates it to be  $1 \mu eV$ by use of tight-binding  and density functional theory ~\cite{min06,yao2007} An all-electron first principles calculation gives a higher value, in order of $50 \mu eV$ \cite{boettger2007}. Regarding d and higher orbitals, the value of ISOC gap is about $24 \mu eV$\cite{gmitra09}. This small ISOC gap results in spin relaxation time of microsecond-order\cite{ertler09,pesin2012}.

 However, much shorter relaxation times in the range of  pico- to few nanoseconds are reported by recent expriments\cite{Tombros07, dlubak2012, drögeler2014, guimaraes2014}.
 This discrepancy between theory and experiment results still remains and the origin of spin relaxation in graphene is a major open question \cite{han2014graphene}.

 The known sources for spin dephasing and scattering are widely discussed in the literature \cite{roche2014,roche2015}. 
   It has been suggested that extrinsic effects~\cite{varykhalov08}, such as substrate and adatoms \cite{ertler09, gmitra2013spin, kochan2014spin}, ripples \cite{huertas09}, and charged impurities \cite{neto09,pi2010}, are the cause of the spin-relaxation time degradation in graphene. 
  
  Substrate surface roughness is another scattering source that can affect spin transport \cite{Ishigami07,lui09,Geringer09}. In an ideal graphene sheet spin-orbit interaction between nearest neighbors vanishes due to the symmetry, whereas next nearest neighbors have negligible effects on spin flipping \cite{kane05}. Surface roughness breaks the symmetry in graphene and enhances spin-orbit interaction between nearest neighbors. Moreover, Substrate surface roughness enhances coupling between  $\mathit{\sigma}$ and $\mathit{\pi}$ orbitals, and therefore spin-orbit interaction.
   In this context, spin transport has been studied for a curved graphene nanoribbon \cite{gosalbez11} and also in the presence of random spin-orbit coupling \cite{huertas07,dugaev11}. However, a careful and comprehensive analysis of the role of surface corrugation on spin-transport is missing. 
  
  In this work, spin transport in corrugated graphene nanoribbon is studied, employing an atomistic approach based on the non-equilibrium Green's function formalism~\cite{PourfathBook14}.

\section{approach}
The Hamiltonian of a graphene nanoribbon (GNR) can be described as: 
\begin{equation}
H = H_\mathrm{TB} + H_\mathrm{SO},
\label{e:H}
\end{equation} 
where $H_\mathrm{TB}$ is the nearest-neighbor tight-binding Hamiltonian excluding spin-orbit coupling: 
\begin{equation}
H_\mathrm{TB}=\sum_{\langle i,j\rangle;l,m} 
(\epsilon_{l,m} \delta_{i,j}\delta_{l,m}-t_{i,j;l,m}) \hat{c}^{\dagger}_{i,j;l,m}\hat{c}_{i,j;l,m},
 \label{eq:htb}
\end{equation} 
where $i$ and $j$ are indices of atoms and $\langle i,j \rangle$ runs over first-nearest neighbors, $l$ and are the orbital indices, which are $p_{x,y,z}$ and $s$ orbitals, $\epsilon$ represents the on-site potential, and $t$ is the hopping parameter. The selected values of discussed parameters can be found in Table~\ref{table:hopping}.

\begin{figure}[!t]
	\centering
	\includegraphics[width=.8\columnwidth]{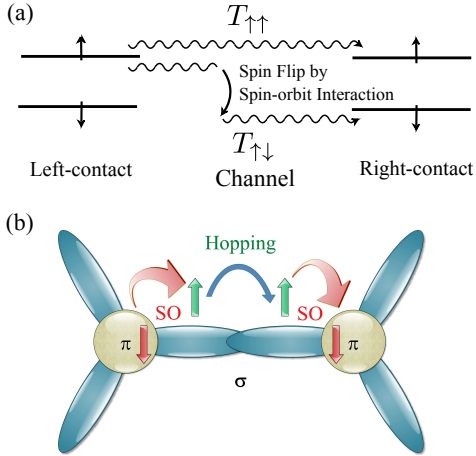}
	\caption{(a) The illustration of spin transport in corrugated graphene nanoribbon. (b) Spin transitions due to spin-orbit interaction in $p_{x,y,z}$ orbitals.}
	\label{f:Schem}
\end{figure}

\begin{table}
	\begin{center}
		\caption{The on-site potentials and hopping parameters of graphene. All values are in the unit of eV}
		\begin{tabular} {p{1.2cm}p{1.2cm}p{1.2cm}p{1.2cm}p{1.2cm}p{1.2cm}   }
			\hline
			\hline
			$\varepsilon_s$  &  $\varepsilon_p$  & $ss\sigma$ & $sp\sigma$ & $pp\sigma$  & $pp\pi$ \\
			\hline
			\hspace{-.3cm}$-7.3$ & $0$ & \hspace{-.3cm}$-4.30$ & \hspace{-.3cm}$+4.98$ & \hspace{-.3cm}$+6.38$ & \hspace{-.3cm}$-2.66$ \\
			\hline 
		\end{tabular}
	\end{center}
	\label{table:hopping}
\end{table}

In Eq.~\ref{e:H} $H_\mathrm{SO}$ is the spin-orbit coupling term that can be written as \cite{Guinea06,gosalbez11}:
\begin{equation}
\begin{split}
& H_\mathrm{SO}=\lambda {\hat{L}\cdot\hat{S}},
\end{split}
 \label{eq:hso}
\end{equation} 
where $\lambda$ is the spin-orbit coupling constant that is chosen to be 12meV. $\hat{L}$ and $\hat{S}$ are angular momentum and spin operators, respectively. Intra-atomic spin-orbit coupling between $p_{x\uparrow}$, $p_{y\uparrow}$, $p_{z\uparrow}$  $p_{x\downarrow}$, $p_{y\downarrow}$, $p_{z\downarrow}$ orbitals is given by:
\begin{equation}
\begin{split}
H_\mathrm{SO}=\frac{\lambda}{2}
& \begin{pmatrix}
  0 & -i & 0 & 0 & 0 & 1\\
  i & 0 & 0 & 0 & 0 & -i\\
  0 & 0 & 0 & -1 & i & 0\\
  0 & 0 & -1 & 0 & i & 0\\
  0 & 0 & -i & -i & 0 & 0\\
  1 & i & 0 & 0 & 0 & 0
    \end{pmatrix}
 \end{split}
\end{equation}

Although s orbital does not contribute to spin-orbit coupling matrix, it affects spin transport through bonding. The ISOC constant is inversely proportional to $\mathit{sp\sigma}$ hopping parameter in second order \cite{min06}.
Substrate surface is an statistical phenomena which can be modeled by a Gaussian auto-correlation function (ACF)~\cite{Ishigami07}:
\begin{equation}
  R(x,y)= \mathit{\delta h}^2\exp\left(-\frac{\mathit{x}^2}{\mathit{L_x}^2}-
  \frac{\mathit{y}^2}{\mathit{L_y}^2} \right)\ .
  \label{eq:corrugation}
\end{equation}

%

$\mathit{L_x}$ and $\mathit{L_y}$ are the roughness correlation lengths along the $x$ and $y$-direction, respectively. $\mathit{\delta h}$ is the root mean square of the height fluctuation. Further details of generating random surface roughness can be found in our previous work \cite{touski13}.
Many samples are generated for one device and then, the characteristics of each device is obtained by an ensemble average over all samples  \cite{yazdanpanah12}. 

\begin{figure}[!t]
	\centering
	\includegraphics[width=1\columnwidth]{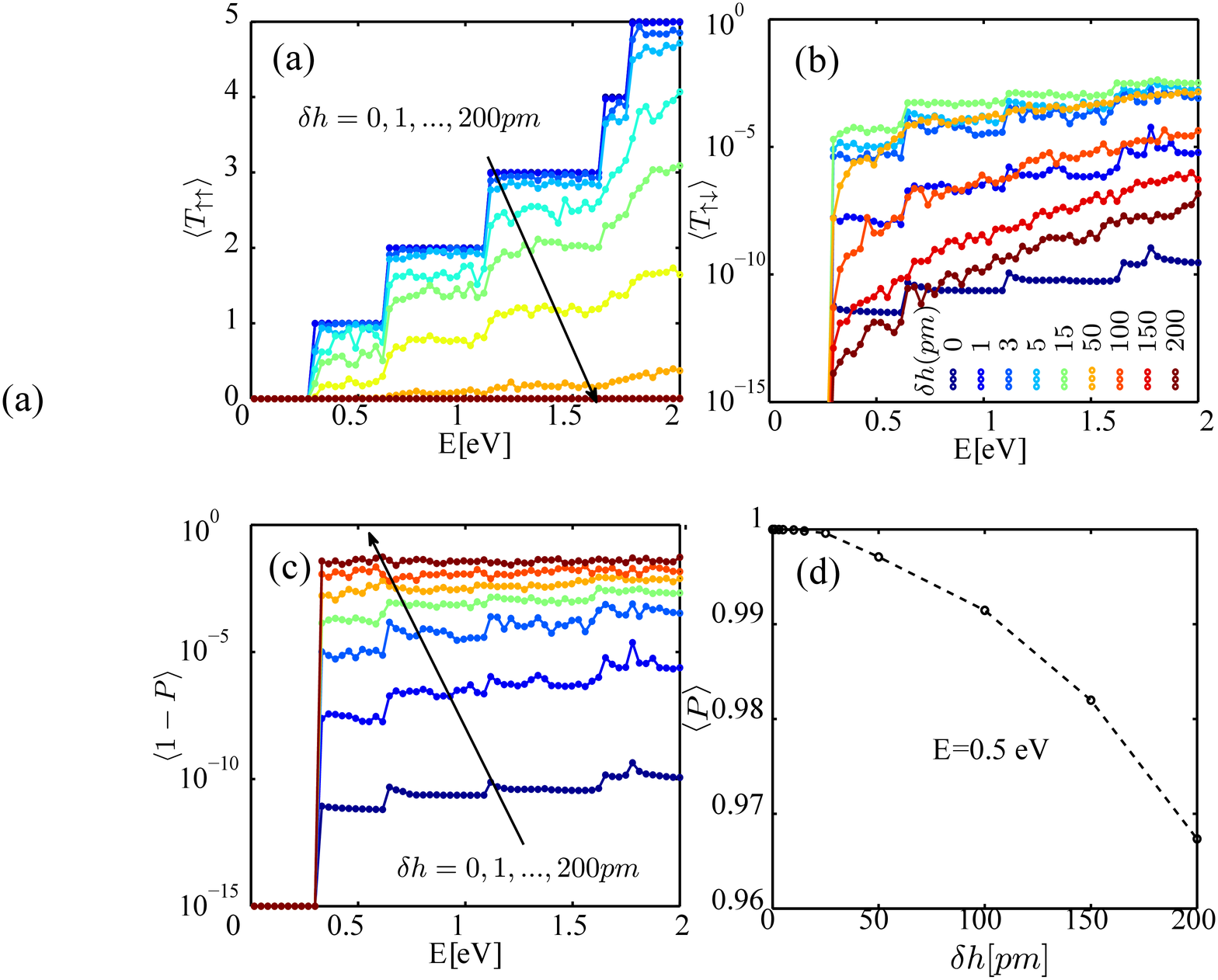}
	\caption{The ensemble average of (a) $T_{\uparrow\uparrow}$, (b) $T_{\uparrow\downarrow}$, and (c) $1-P$ as a function of energy at various substrate surface roughness amplitudes in AGNR with $NW=15$, $L=40nm$ and $L_x,L_y=10nm$ and (d) $P$ versus substrate surface roughness amplitude at $0.5eV$.}
	\label{f:H}\label{f:H}
\end{figure}

\begin{table}
	\begin{center}
		\caption{Surface rughness parameters of some materials \cite{lui2009,ishigami2007, geringer2009,dean2010,kwon2009,wang2008}}
		\begin{tabular} {c c c c c }
			\hline
			\hline
			&  SiO$_2$  & BN & Mica & Al$_{2}$O$_{3}$\\
			\hline
			$\delta h$ & $\mathrm{168-360pm}$ & $\mathrm{\simeq 75pm}$ &$\mathrm{24pm}$ & $\mathrm{330pm}$\\
			\hline 
			$L_x=L_y$ & $\mathrm{15-32nm}$ & - & $\mathrm{2nm}$ & -\\
			\hline 
		\end{tabular}
	\end{center}
	\label{table:rough}
\end{table}

The amplitude of Graphene surface roughness is depend on its substrate material and polishing method(see Table~\ref{table:rough}).
 Surface roughness alters the positions and directions of atomic orbitals of graphene. The hopping parameters have been modulated using Harrison's model $t_{ij}\propto 1/d^2$ \cite{harrison99} and the effect of orbital direction variation on the hoping parameters has been included by Slater-Koster model \cite{slater03}.

The non-equilibrium Green's function is used to study the spin transport in corrugated nanoribbons. The Green's function of the channel is given by:
\begin{equation}
\begin{split}
& \begin{pmatrix}
  G^{r,a}_{\uparrow\uparrow}(\epsilon)    & G^{r,a}_{\uparrow\downarrow}(\epsilon) \\
  G^{r,a}_{\downarrow\uparrow}(\epsilon)  & G^{r,a}_{\downarrow\downarrow}(\epsilon) 
 \end{pmatrix}
 =  \left[ (\epsilon \pm i\eta)I -
 \begin{pmatrix}
  H_{\uparrow\uparrow}    & H_{\uparrow\downarrow} \\
  H_{\downarrow\uparrow}  & H_{\downarrow\downarrow}
 \end{pmatrix}
 \right. \\
& \left. ~~~~ 
  - 
   \begin{pmatrix}
  \Sigma_{L\uparrow}    & 0 \\
  0                               & \Sigma_{L\downarrow}
 \end{pmatrix}
   - \begin{pmatrix}
  \Sigma_{R\uparrow}    & 0 \\
  0                               & \Sigma_{R\downarrow}
 \end{pmatrix}
  \right]^{-1},
  \end{split}
\end{equation}
where $\eta$ is an infinitesimal quantity and $\Sigma_{R/L,\sigma}$ is the self energy of the left or right contact for electrons with up or down spin ($\sigma=\uparrow,\downarrow$) that is given by
\begin{equation}
\Sigma_{\alpha,\sigma} = \tau_{\alpha,\sigma}^{\dagger} g_{\alpha,\sigma}(\epsilon) \tau_{\alpha,\sigma}.
\end{equation} 
$g_{\alpha}$ is the surface Green's function of semi-infinite left and right electrodes which can be obtained from highly convergent Sancho method \cite{sancho84}. The transmission probability can be obtained by \cite{wang2001nonlinear}:

\begin{subequations}
\begin{align}
& T_{\uparrow\uparrow}=\mathrm{Trace}[\Gamma_{L\uparrow}G^r_{\uparrow\uparrow}\Gamma_{R\uparrow}
G^a_{\uparrow\uparrow}],\\
& T_{\uparrow\downarrow}=\mathrm{Trace}[\Gamma_{L\uparrow}G^r_{\uparrow\downarrow}\Gamma_{R\downarrow}
G^a_{\downarrow\uparrow}] \,
\end{align}
\label{s:Tsigma}
\end{subequations}

\begin{figure}[!t]
	\centering
	\includegraphics[width=1\columnwidth]{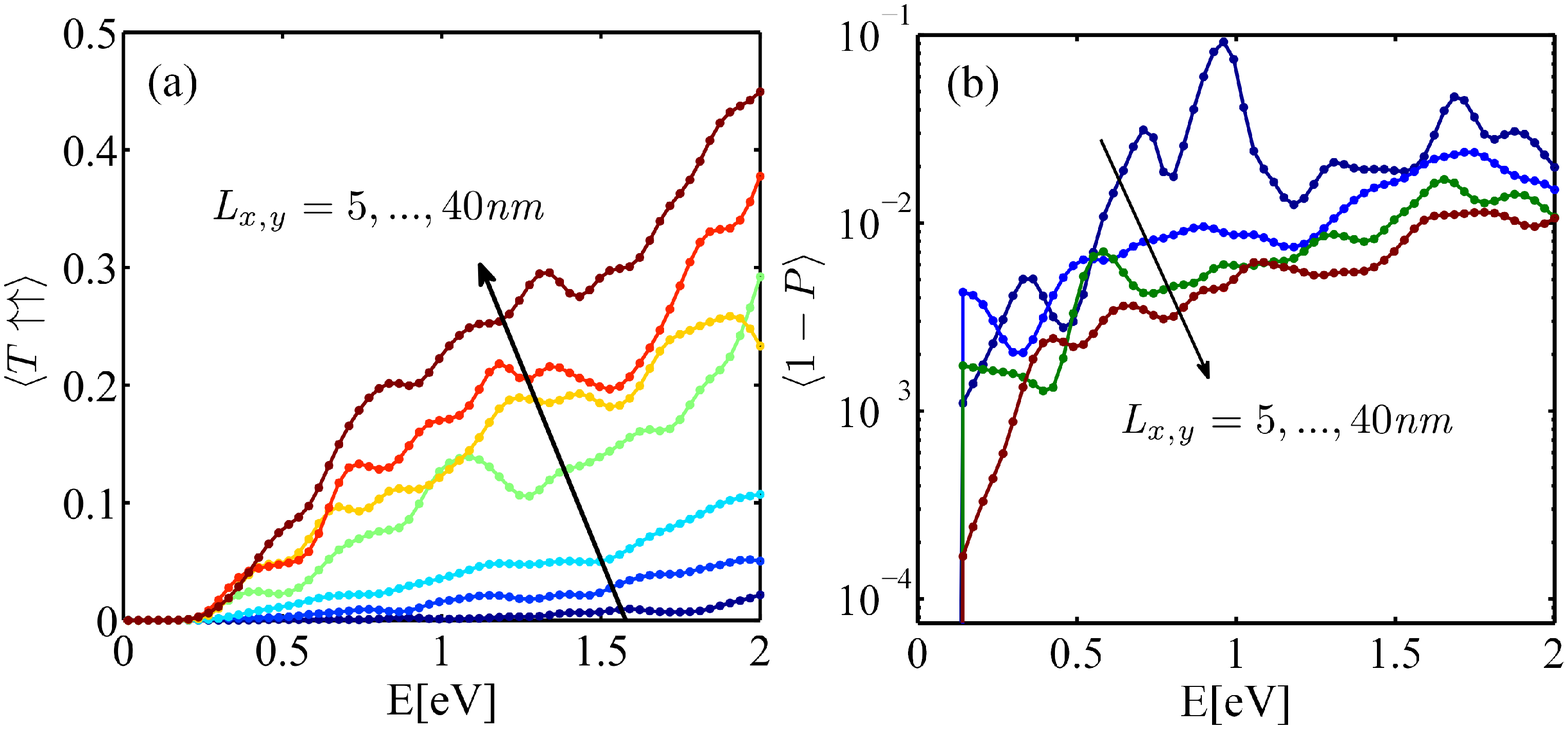}
	\caption{(a) $T_{\uparrow\uparrow}$ and (b) polarization as a functions of energy at various roughness correlation lengths for AGNR with $NW=15$, $L=40nm$ and $\delta h=100pm$.}
	\label{f:TElx}
\end{figure}


 $T_{\uparrow\uparrow}$ denotes the transmission probability of carriers with up-spin in the left contact to up-spin in the right-contact, and $T_{\uparrow\downarrow}$ refers transmission probability of carriers with up-spin in the left-contact to down-spin in the right-contact as shown in the Fig.\ref{f:Schem}. $T_{\uparrow\downarrow}$ is an indication of spin-flip along the channel due to spin-orbit interaction. Corrugation and spin-orbit interaction are neglected in the contacts. $\Gamma_{\alpha\sigma}$ is the broadening of the left and right-contact for up- and down-spin, which is defined as:
 \begin{equation}
\Gamma_{\alpha,\sigma} = i(\Sigma_{\alpha,\sigma}(\epsilon) - \Sigma_{\alpha,\sigma}^{\dagger}(\epsilon)).
\end{equation}

\section{results and discussion}

In this work an AGNR channel is considered. Surface roughness is applied and spin transport in the device is studied.

$T_{\uparrow\uparrow}$ and $T_{\uparrow\downarrow}$ are plotted as a function of energy at various roughness amplitudes in the \Fig{f:H}. $T_{\uparrow\uparrow}$ decreases as roughness amplitude increases due to scattering with surface and increasing spin-orbit interaction (spin flip), while the variation of $T_{\uparrow\downarrow}$ is more complicated. One expects that due to enhanced spin-orbit interaction and the resulting spin flip rate, $T_{\uparrow\downarrow}$ should increase with roughness amplitude. However, $T_{\uparrow\downarrow}$ shows a descending behavior with roughness amplitude. This behavior can be explained by the fact that surface roughness causes two scattering processes: (I) spin flipping due to enhanced spin-orbit interaction and (II) scattering due to random variation of hopping parameters. At small roughness values the former results in $T_{\uparrow\downarrow}$ increment, whereas at larger roughness values the latter dominates and results in the reduction of $T_{\uparrow\downarrow}$. To focus on spin-flipping process spin-polarization can be defined as \cite{Perel03}:
\begin{equation}
 P=\frac{T_{\uparrow\uparrow} - T_{\uparrow\downarrow}}{T_{\uparrow\uparrow} + T_{\uparrow\downarrow}} ,
\end{equation} 
which is spin-polarization of carriers through the channel. As shown in the \Fig{f:H}, (c) and (d) $P$ decreases as roughness amplitude increases. $1-\langle P \rangle = 2T_{\uparrow\downarrow}/(T_{\uparrow\uparrow}+T_{\uparrow\downarrow})$ indicates spin flip transmitted over total transmission.

\begin{figure}[!t]
	\centering
	\includegraphics[width=1\columnwidth]{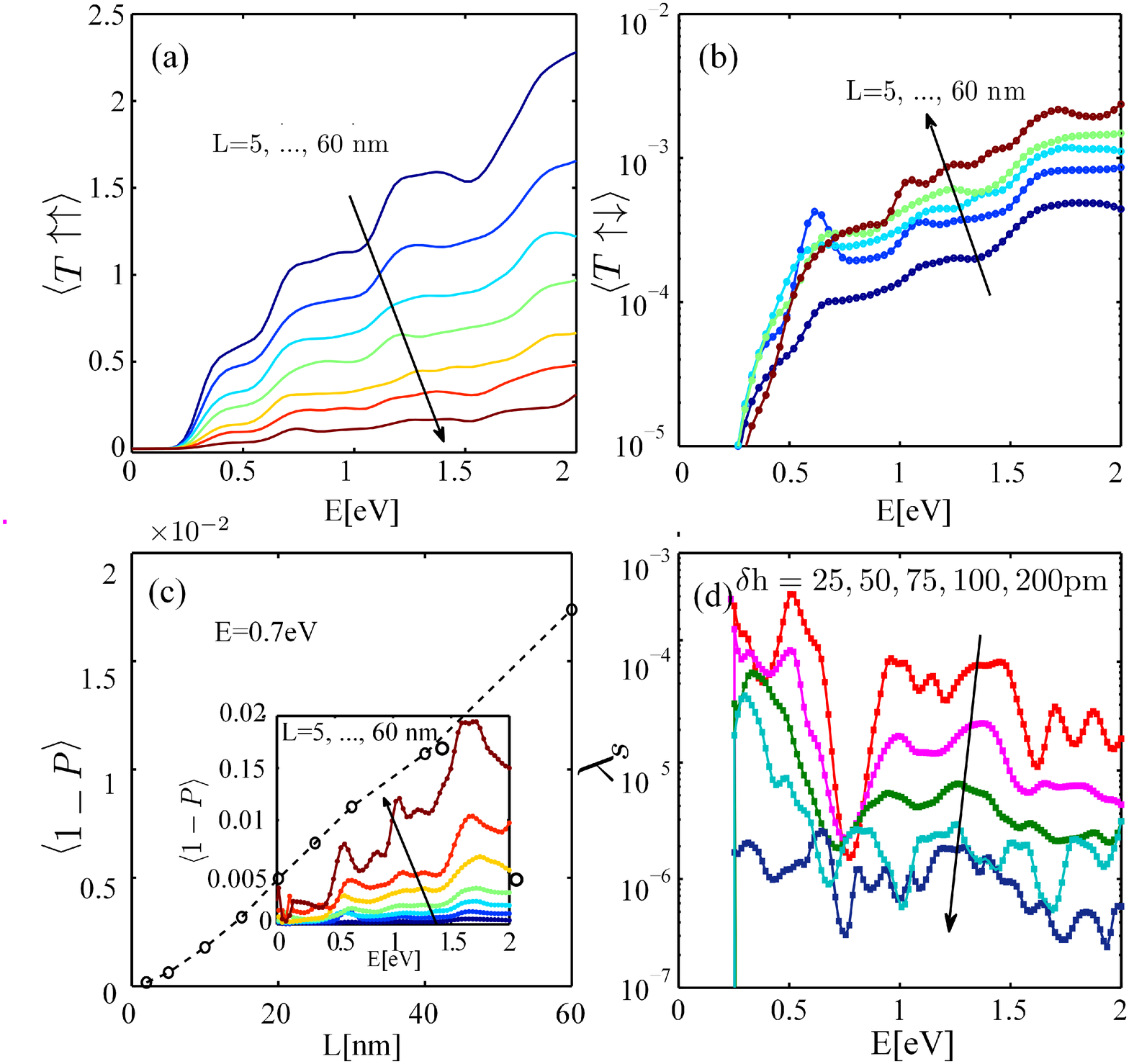}
	\caption{The ensemble averages of (a) $T_{\uparrow\uparrow}$, (b) $T_{\uparrow\downarrow}$ as functions of energy at various channel lengths, (c) the polarization versus channel length for $0.7eV$, $NW=15$, $\delta h=50pm$ and $L_x,L_y=10nm$, and (d) spin-diffusion length $(\lambda_s)$ at various roughness amplitudes.}
	\label{f:L}
\end{figure}

The results, in accordance to previous works\cite{huertas2006,chico2009,lopez2011}, indicate increasing spin-flip rate as a consequence of surface roughness increasing. This is because the local curvature mixes $\pi$ and $\sigma$ bonding. 

Correlation length has a different effect on transmission probability, see \Fig{f:TElx}. $T_{\uparrow\uparrow}$ increases with correlation length. A longer correlation length leads to a smoother substrate surface that results in the reduction of scattering and spin-flip rate. Therefore, both $T_{\uparrow\uparrow}$ and polarization increase with correlation length.

Spin flipping occurs during carrier transport through the channel. The effects of the channel length on the spin transmission are investigated in \Fig{f:L}. $T_{\uparrow\uparrow}$ decreases as the channel length increases due to increased scattering and spin-flipping, while  $T_{\uparrow\downarrow}$ increases with the channel length. The polarization versus energy at various channel lengths is shown in \Fig{f:L}. To quantify the effect of surface corrugation on the spin-polarization  one can define spin diffusion length $(\lambda_s)$ as $P(\epsilon)=P_0\exp(-L/\lambda_s(\epsilon))$. To extract spin diffusion length the polarization as a function of channel length is plotted  (\Fig{f:L} (c)) and an exponential curve is fitted and the respective $\lambda_s$ is evaluated.

\begin{table}[!t]
	\begin{center}
		\caption{Spin diffusion length for $\mathit{L_x,L_y=10}\mathrm{pm}$ and various $\delta h$ .}
		\begin{tabular} {l c c c c c c}
			\hline
			\hline
			$\delta h$  & $\mathrm{25pm}$ & $\mathrm{50pm}$ & $\mathrm{75pm}$ & $\mathrm{100pm}$ & $\mathrm{150pm}$ & $\mathrm{200pm}$\\
			\hline
			$L_s$ & $\mathrm{80\mu m}$ & $\mathrm{29\mu m}$ & $\mathrm{12\mu m}$ & $\mathrm{5\mu m}$ & $\mathrm{1.6\mu m}$ & $\mathrm{1.1\mu m}$\\
			\hline 
			
		\end{tabular}
	\end{center}
	\label{table:sdl}
\end{table}

 
Spin diffusion length for surface roughness amplitude of $\mathrm{200 pm}$, witch is common for graphene on $\mathrm{SiO_2}$ substrate, is obtained by averaging over spin diffusion length in various energy. the value of spin diffusion length is equal to $\mathrm{1.2 \mu m} $.

The $L_s$ for various surface roughness amplitude can be found in Table~\ref{table:hopping}.
%

 
  The spin relaxation time can be obtained from $\tau_s=\lambda^2/D_s$, with a spin diffusion coefficient of $D_s\approx 10^{-2}m^2/s$ \cite{han11}. The extracted spin relaxation time is in order of picosecond which is in agreement with experimental results \cite{han11,MHD12}.

\section{conclusions}
The effects of surface roughness on spin transport in AGNR is investigated. The influence of roughness parameters and channel length  is studied. It is shown that surface roughness has a significant influence on spin-polarization. By increasing roughness amplitude the spin-polarization decreases due to increased spin-flip induced by symmetry breaking and entanglement of $\mathit{\sigma}$ and $\mathit{\pi}$ orbitals, with surface roughness. However, the increase of the correlation length leads to conservation of spin-polarization over longer channel lengths. The effect of channel length is studied and the spin-relaxation time is obtained in the order of picosecond  which is agreement with experimental measurements which  indicates the importance of including surface roughness for understanding the relatively small spin-relaxation time in graphene.



\end{document}